\documentclass[fleqn,12pt,twoside]{article}
\usepackage{espcrc1}

\usepackage{graphicx}
\usepackage{epsf,epsfig,psfig,float,amssymb,stmaryrd,latexsym}
\usepackage{graphics,psfrag}
\usepackage[figuresright]{rotating}

\def\3{{\ss}}

\def\vek #1 {\overrightarrow {#1}}

\newcommand{\vs}{\vspace{-0.0cm}}
\newcommand{\beq}{\begin{equation}}
\newcommand{\eeq}{\end{equation}}
\newcommand{\beqn}{\begin{displaymath}}
\newcommand{\eeqn}{\end{displaymath}}
\newcommand{\beqa}{\begin{eqnarray}}
\newcommand{\eeqa}{\end{eqnarray}}
\newcommand{\beqan}{\begin{eqnarray*}}
\newcommand{\eeqan}{\end{eqnarray*}}

\newcommand{\bma}{\begin{array}{cc}}
\newcommand{\ema}{\end{array}}

\newcommand{\AmS}{{\protect\the\textfont2
  A\kern-.1667em\lower.5ex\hbox{M}\kern-.125emS}}

\title{The chiral limit of QCD and above}

\author{Ulf-G. Mei{\ss}ner\address{Universit\"at Bonn, HISKP,
             D-53115 Bonn, Germany}
  \address{Forschungszentrum J\"ulich, IKP (Th),
  D-52425 J\"ulich, Germany }}

\begin{document}

\maketitle

\begin{abstract}
I review aspects of chiral dynamics pertinent to the structure of baryons and
few-nucleon systems, such as chiral extrapolations for the nucleon and the
delta mass, double pion photoproduction off protons, single neutral pion 
electroproduction off the deuteron, pion photoproduction in the delta
region, the quark mass dependence of the nuclear
forces and the possibility of an infrared limit cycle in QCD.
\end{abstract}

\section{CHIRAL LIMIT OF QCD AND EFFECTIVE FIELD THEORY}
 
In QCD, the up, down and strange quarks are light compared to the typical
hadronic scale. Therefore, it is a good first approximation to consider the
SU(3) chiral limit with $m_u=m_d=m_s=0$ (and all heavy quarks 
decoupled, $m_c=m_b=m_t =\infty$). 
In that limit, QCD posseses an exact SU(3)$_L \times$SU(3)$_R$ chiral symmetry
and has a high degree of symmetry because gluons are flavor-blind. However,
the full symmetry is not shared by the vacuum, the chiral symmetry is spontaneously
broken with the appearance of eight massless pseudoscalar mesons, to be 
identified with the pions, the kaons and the eta. These mesons are pseudo-Goldstone
bosons because of the explicit symmetry breaking through the quark masses, 
collected in the quark mass matrix ${\cal M}$. The
consequences of the explicit and the spontaneous symmetry breaking can be
analyzed utilizing an Effective Field Theory (EFT), Chiral Perturbation Theory
(CHPT), or variants thereof. For a review, see e.g. \cite{BKMrev}. For the
following discussion, I briefly recapitulate some salient features of the
chiral limit (CL) of QCD. While S-matrix elements exist in the CL for
arbitrary momenta, the approach to the CL is non-analytic in ${\cal M}$,
which leads to the famous ``chiral logs'' as pointed out by many. It is important
that there further exists the so-called decoupling theorem \cite{GZ}:
Leading chiral non-analytic terms stem from pion (Goldstone boson) one-loop 
graphs coupled to pions (Goldstone bosons) or nucleons (ground state baryons).
This means that resonances like the $\rho$ or the $\Delta$ decouple, and
this constraint must be implemented when one constructs EFTs with explicit
resonance degrees of freedom. Furthermore, the chiral limit of QCD and
the large $N_c$ limit do not commute, which is another constraint when setting
up EFTs. More on that below. I will now discuss various issues related to the
EFT of QCD that are pertinent to the structure of baryons. 

\section{QUARK MASS DEPENDENCE OF NUCLEON AND DELTA MASSES}

\begin{figure}[tb]
\hfill
\psfig{file=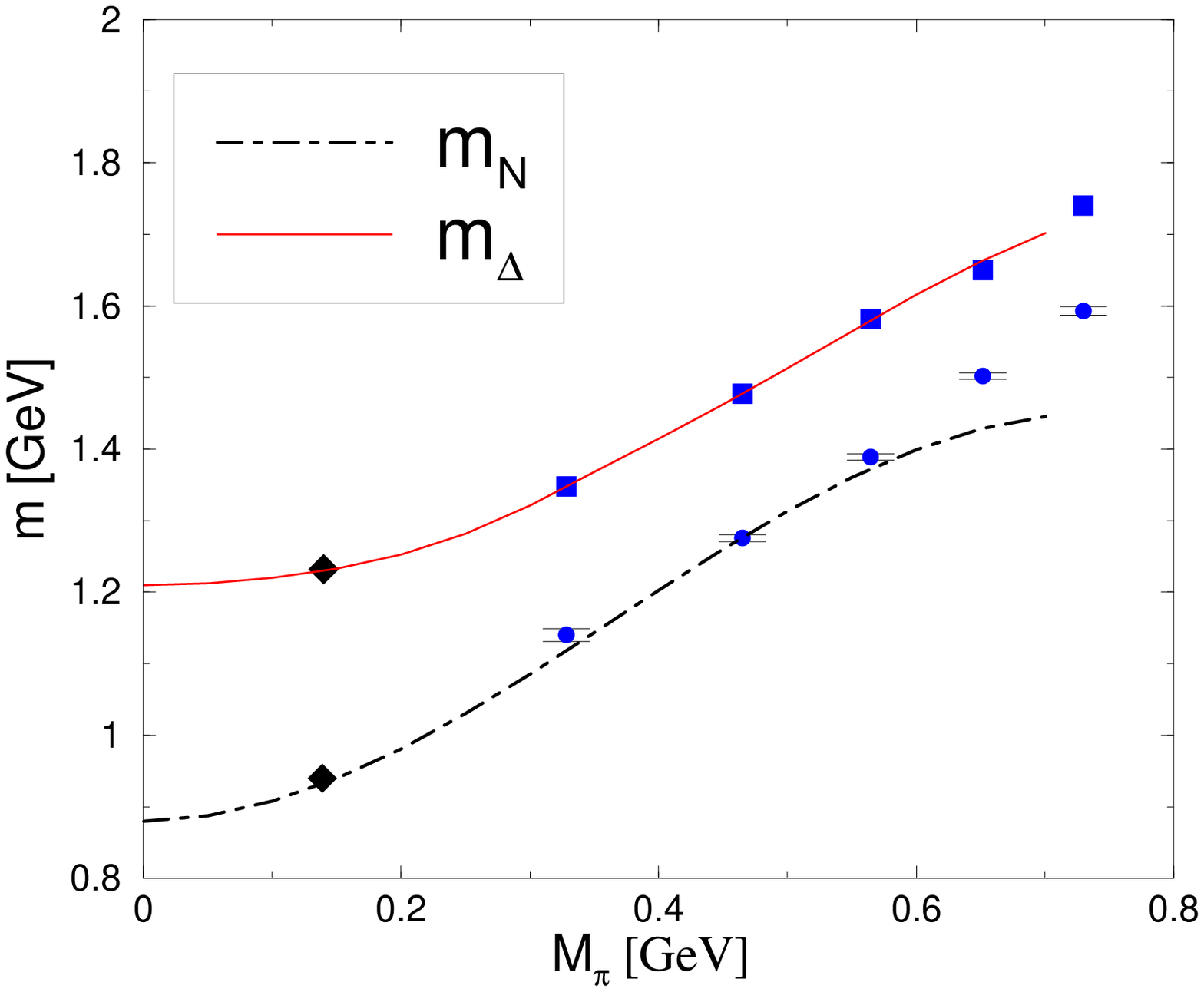,width=6.9cm}
\vspace{-7.4cm}
\flushleft{
\psfig{file=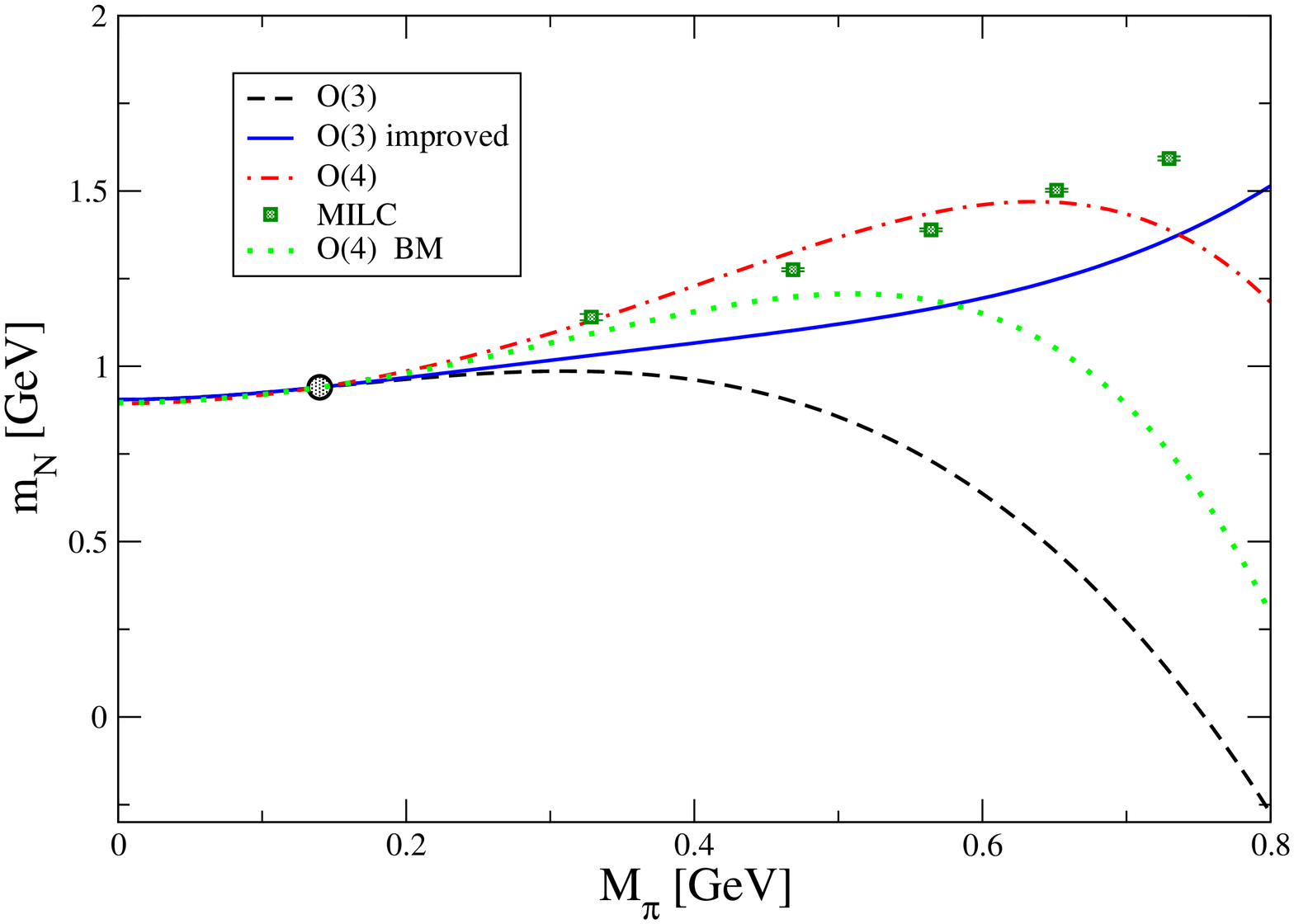,width=6.5cm,angle=270}
}

\vskip -0.7 true cm
\centerline{
\parbox{1.00\textwidth}{
\caption[fig1]{
\label{fig1}  
Left panel: Pion mass dependence of the nucleon mass in an SU(3) calculation 
at third (dashed), improved third (solid) and fourth (dot-dashed) order, 
respectively. The dotted line represents the fourth order calculation 
from \protect\cite{BM}.
Right panel: Pion mass dependence of $m_N$ (dot-dashed) and $m_\Delta$ 
(solid line) based on a covariant 
calculation in the small scale expansion. The filled diamonds denote the
physical values of $m_N$ and $m_\Delta$ at $M_\pi =140\,$MeV.
The data in both panels are from MILC
\protect\cite{MILC}.
}}}
\end{figure}

CHPT can be used to connect the results of lattice simulations at unphysical
quark masses with the world at their physical values. Such procedures are frequently
denoted as ``chiral extrapolations''. Of course, as the quark masses increase,
the pion (and also the kaon) becomes heavier, eventually rendering such a
procedure meaningless. More lattice results at lower quark masses are urgently
called for, but it is very instructive to set up the framework and study how
far one can get. Since this topic is still in its developing phase, I will not try to cover
all possible angles but rather concentrate on some recent work concerning the
nucleon, the delta and also the ground state octet baryons. For other views, the
reader is refered to the contributions of Leinweber \cite{Aus} and Procura
\cite{TUM} to this conference. The nucleon mass and general features of the
nucleon were studied in \cite{BHMcut} utilizing dimensional (DR) 
and cut-off (CR) regularization, working to  fourth order in the chiral expansion,
\begin{equation}
m_N = m_0  - 4c_1 M_\pi^2 - {\frac{3g_A^2 M_\pi^3}{32\pi F_\pi^2}} +
k_1 \, M_\pi^4 \ln {M_\pi \over m_N} + k_2\, M_\pi^4 +
{\cal O}(M_\pi^5)~,
\end{equation}
with $m_0$ the nucleon mass in the {\em chiral SU(2)} limit ($m_u=m_d=0$,
$m_s$ fixed at its physical value),
$M_\pi$ ($F_\pi$) the pion mass (decay constant), and $c_1,
k_1, k_2$ are (combinations of) second and fourth order LECs that can
e.g. be determined in the CHPT analysis of pion-nucleon scattering, see
\cite{piN}. With this input, one finds $m_0 = 880\,$MeV and a large theoretical
uncertainty for pion masses larger than $\sim 550\,$MeV. This work was
recently extended to the three-flavor case \cite{FMS}, cf. the left panel
in Fig.~\ref{fig1}. As shown in that figure, a slight readjustment of the
fourth order LECs earlier determined in \cite{BM} allows quite well to describe
the trend of the (partially quenched) data from the MILC collaboration
\cite{MILC}, compare the dot-dashed (new LECs) and the dotted (old LECs)
line. Also shown in that
figure are the result at third order (dashed) and including an improvement
term (solid line) as suggested in \cite{BHMcut}. Clearly, one has to work
at fourth order if one wants to describe the lattice data below pion masses
of about 600~MeV (the curves are only shown for larger $M_\pi$ to better
exhibit the trends). 
%
%
\noindent
The pion mass dependence of the nucleon mass is very similar
to what is found in the two-flavor case. The kaon mass dependence of $m_N$ is also
studied \cite{FMS}; it is less precisely determined. One finds e.g. for the nucleon
mass in the {\em chiral SU(3)} limit the wide range 
$m_0^{SU(3)} \in [710,1070]\,$MeV. In that paper, for the first time results
for the $\Lambda$, the $\Sigma$ and the $\Xi$ were given and compared to
the few existing lattice date.
These results are encouraging and the full machinery of
partially quenched CHPT (see e.g. \cite{AWL}) should be applied to the MILC data.
To make similar statements for the delta resonance, one has to include it
in the EFT explicitly. This can be done by counting the $N\Delta$ mass
splitting as an additional parameter - which, however, does not vanish in the
chiral limit. This is a nice example of the decoupling theorem in the CL and the
resulting non-commutativity with the large $N_C$ limit. The generalized
(phenomenological) power
counting including $m_\Delta -m_N$ is called the ``small scale expansion'' 
(SSE) \cite{HHK}. The quark mass dependence of the delta mass based on a covariant 
version of the SSE \cite{BHMdelta} was recently worked out and the central
results of this study are shown in the right panel of Fig.~\ref{fig1}. The
effective $\pi N \Delta$ Lagrangian includes additional parameters that allow
for a simultaneous description of the nucleon and the delta mass. It is interesting
that the lattice data shown in the figure (again from MILC) 
seem to indicate a stronger slope
of  $m_N (M_\pi)$ than for $m_\Delta (M_\pi)$ for pion masses slightly above
the physical point. This indicates that the pion-delta sigma term is sizeably
smaller than the pion-nucleon one - in the strict SU(6) limit one would expect
$\sigma_{\pi \Delta} = \sigma_{\pi N}$. This indicates that pion cloud effects
are less pronounced in the baryon resonances than in the ground-state. Note that
the MILC data shown in the figure are again from partially quenched SU(3) 
simulations - thus it would be interesting to repeat this analysis in the
framework presented in \cite{TWL}.


\section{PHOTO-NUCLEON/NUCLEAR PROCESSES}

Pion photo- and electroproduction has been established as one of the major
testing grounds of baryon chiral dynamics. Shortly after the detailed investigations
of single neutral pion production off nucleons (see e.g. the review \cite{BKMrev}), 
electromagnetic two-pion production off
protons and neutrons was also considered. Naively, one would estimate the
cross section to be very small, more precisely $\sigma_{\rm tot} 
(\gamma p \to \pi^0 \pi^0 p) \ll \sigma_{\rm tot} (\gamma p \to \pi^+ \pi^- p)$
because of the ``double Kroll-Rudermann suppression''. In the pioneering paper
in 1994 \cite{BKMS2pi} it was, however, shown, that a) there are large chiral 
loop corrections in the $2\pi^0$ channel and b) that the leading 
$\Delta$-contributions cancel. At third order and in the threshold region, 
one finds indeed that $\sigma_{\rm tot}(\gamma p \to \pi^0 \pi^0 p) >
\sigma_{\rm tot} (\gamma p \to \pi^+ \pi^- p)$. This was sharpened a few years
later when the calculation was extended to fourth order and the cross section
for $\gamma p \to \pi^0 \pi^0 p$ could be given in analytic form \cite{BKM2pi}:
\begin{equation}\label{2pi}
\sigma_{\rm tot}(E_\gamma) = {\mathcal C} \, [{\rm nb}] \, \biggl(
{E_\gamma - E_\gamma^{\rm thr} \over 10\, {\rm MeV}}
 \biggr)^2 \,, \quad {\mathcal C} =
\left\{ \begin{array}{ll} 0.6 & {\rm central~value~,} \\
                          0.9 & {\rm upper~limit~,} \end{array} \right. 
\end{equation}
Here, $E_\gamma^{\rm thr} = 308.8\, {\rm MeV}$ is the threshold energy and the
constant ${\mathcal C}$ contains some low-energy constants, in particular one
related to the decay of the Roper into two pions,
$N^* (1440) \to N (\pi\pi)_{\rm S-wave}$. From the earlier study of the
reaction $\pi N \to \pi\pi N$ in CHPT a central value as well as an upper limit for
this constant could be given \cite{BKMpipiN} 
which reflects itself in the values for  ${\mathcal C}$ in Eq.~(\ref{2pi}).
\begin{figure}[tb]
\begin{minipage}{6cm}
Figure 2.  Total cross section for two-neutral-pion 
photoproduction off the proton, $\gamma p \to 
\pi^0\pi^0 p$, in the
threshold region as a function of the photon
energy $E_\gamma$. 
The solid line is the fourth order
result of \protect\cite{BKM2pi} and the dashed line
refers to the upper limit also given in that paper.
The data are from the recent measurement of the TAPS
collaboration \cite{TAPS}. The threshold energy is
$E_\gamma^{\rm thr} = 308.8\,$MeV.
\end{minipage}
\hskip 1.7 true cm
\begin{minipage}{8cm}
\psfig{file=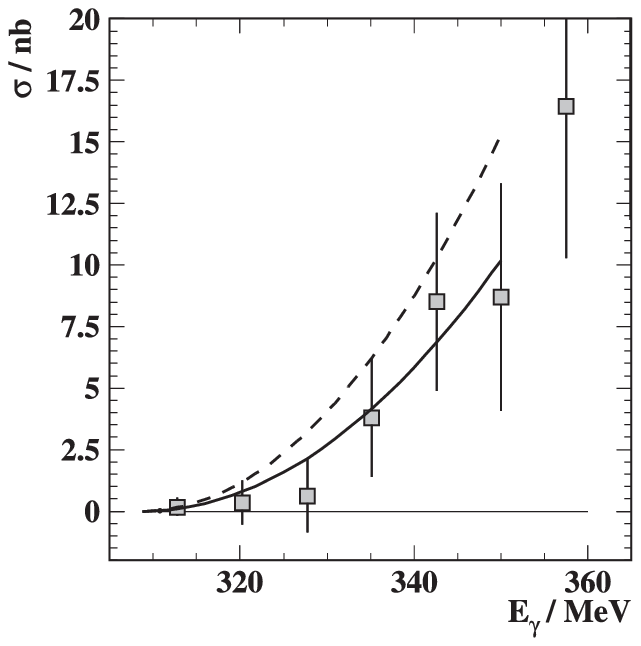,width=7.5cm}
\end{minipage}
\label{fig2}
\end{figure}
\setcounter{figure}{2}
\noindent This sharp prediction of baryon CHPT could only be tested many years
later because the predicted cross section in the threshold region is very small
and a dedicated experiment with the TAPS detector at MAMI had to be performed.
The TAPS collaboration published their result in 2004 and it agrees beautifully
with the central prediction, cf. Fig.~\ref{fig2}. They even state that the
upper limit for the $N^* (1440) \to N (\pi\pi)_{\rm S-wave}$ can be excluded. 
The experimental result is also in agreement with a result obtained in the chiral
unitary model of the Valencia group \cite{Roca}, where the pion loop effects
were generated by pion rescattering in the scalar-isoscalar channel.

\begin{figure}[tb]
\centerline{
\psfig{file=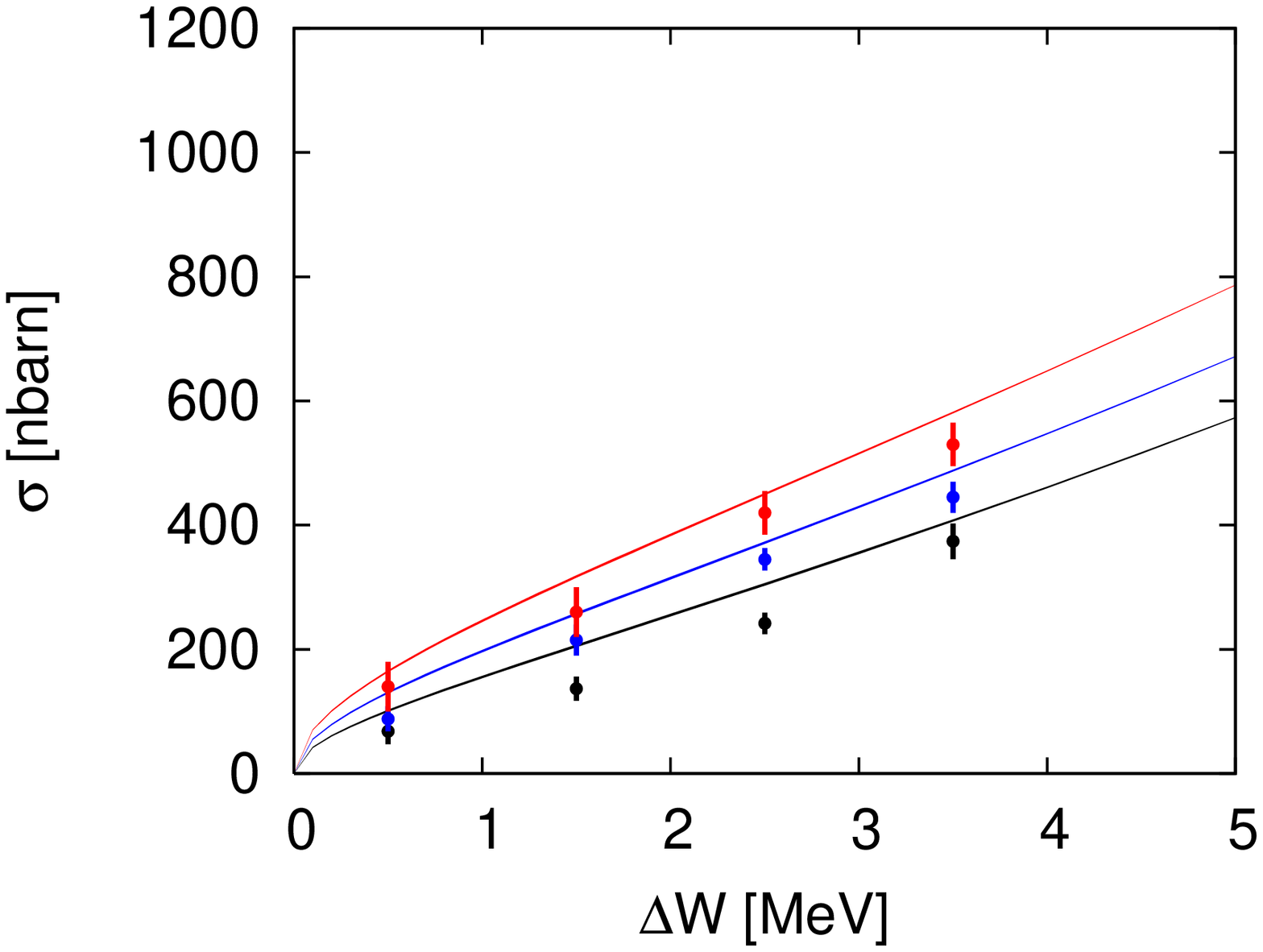,width=6.2cm,angle=270}
\psfig{file=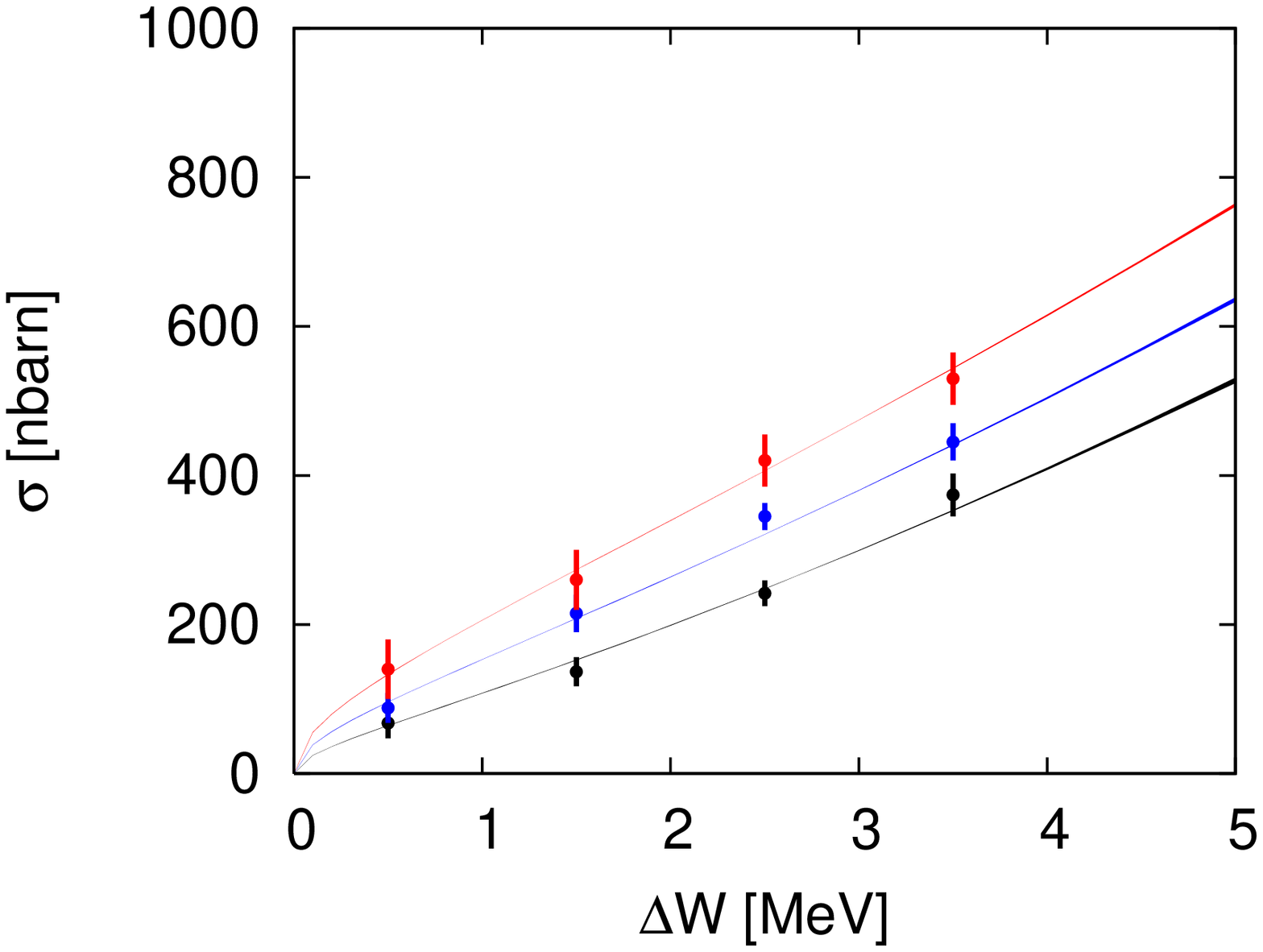,width=6.2cm,angle=270}
}
\vskip -0.7 true cm
\centerline{
\parbox{1.00\textwidth}{
\caption[fig3]{
\label{fig3}  
   Left panel: Total cross section as a function of $\Delta W$ for three 
   different values of the photon polarization in comparison 
   to the MAMI data \protect\cite{Ewald} for fit~1 and the 
   NNLO wave functions. $\Delta W$ is the pion  energy above threshold. 
   The upper/middle/lower band corresponds to the largest/medium/smallest 
   value of $\varepsilon$. 
   Right panel: Same for fit~2.
}}}
\vspace{0.05cm}
\end{figure}

Pion photo- and electroproduction off the deuteron allows to test the
counterintuitive CHPT prediction that the S-wave amplitude $E_{0+} (\pi^0 n)$
is larger in magnitude than the one for the proton, $E_{0+} (\pi^0 p)$
\cite{BKMZ}. In
a naive dipole picture, one expects $E_{0+} (\pi^0 n) = 0$. In the case of
photoproduction, only one measurement of the total cross section with sufficient
accuracy has ever been performed at the now defunct Saskatoon accelerator SAL
\cite{SAL}. The extrapolated threshold cross section agrees with the earlier
CHPT prediction \cite{BBLMvK} and clearly rules out a vanishing neutron dipole
amplitude. At that time, the hybrid approach was used, i.e. the deuteron
wave functions were taken from precise phenomenological potentials and then
applied to the kernel calculated within CHPT. 
More data have become available with the MAMI measurements of coherent
pion electroproduction, $\gamma (Q^2) d \to \pi^0 d$, at the 
(negative of the) photon virtuality
$Q^2 = 0.1\,$GeV$^2$ \cite{Ewald}. This reaction has been analyzed in detail
in Refs.\cite{KBM}. In particular, the threshold multipole expansion has been
developed, consistent deuteron wave functions from next-to-next-to-leading
order nuclear chiral EFT \cite{EGMZ} were utilized and boost effects were 
considered. The pertinent matrix elements
decompose into the so-called single scattering and the three-body terms. While
the former contain the information on the elementary proton and the neutron
amplitudes, the latter are parameter-free at third order.  The fourth order
three-body corrections contain two four-nucleon LECs. In the fits shown in
Fig.~\ref{fig3}, these were considered as free parameters. The two fits shown in 
that figure represent the theoretical uncertainty at that order. In fit~1 
(left panel), one fits to the longitudinal deuteron S-wave multipole $|L_d|$
at $Q^2 = 0.1\,$GeV$^2$ as deduced in \cite{Ewald} whereas for fit~2 a best
description of the total cross section data is achieved. One  notices
that the uncertainty due to the variation in the chiral wave functions is very
small, the various lines are indeed bands that cover the set of wave functions
given in \cite{EGMZ}. Again, these data are clearly indicative of  non-vanishing
longitudinal and transverse neutron dipole amplitudes. Also, the resulting 
deuteron electric dipole amplitude at
the photon point $Q^2 = 0$ is consistent with the SAL result \cite{SAL}.

\begin{figure}[htb]
\centerline{
\psfig{file=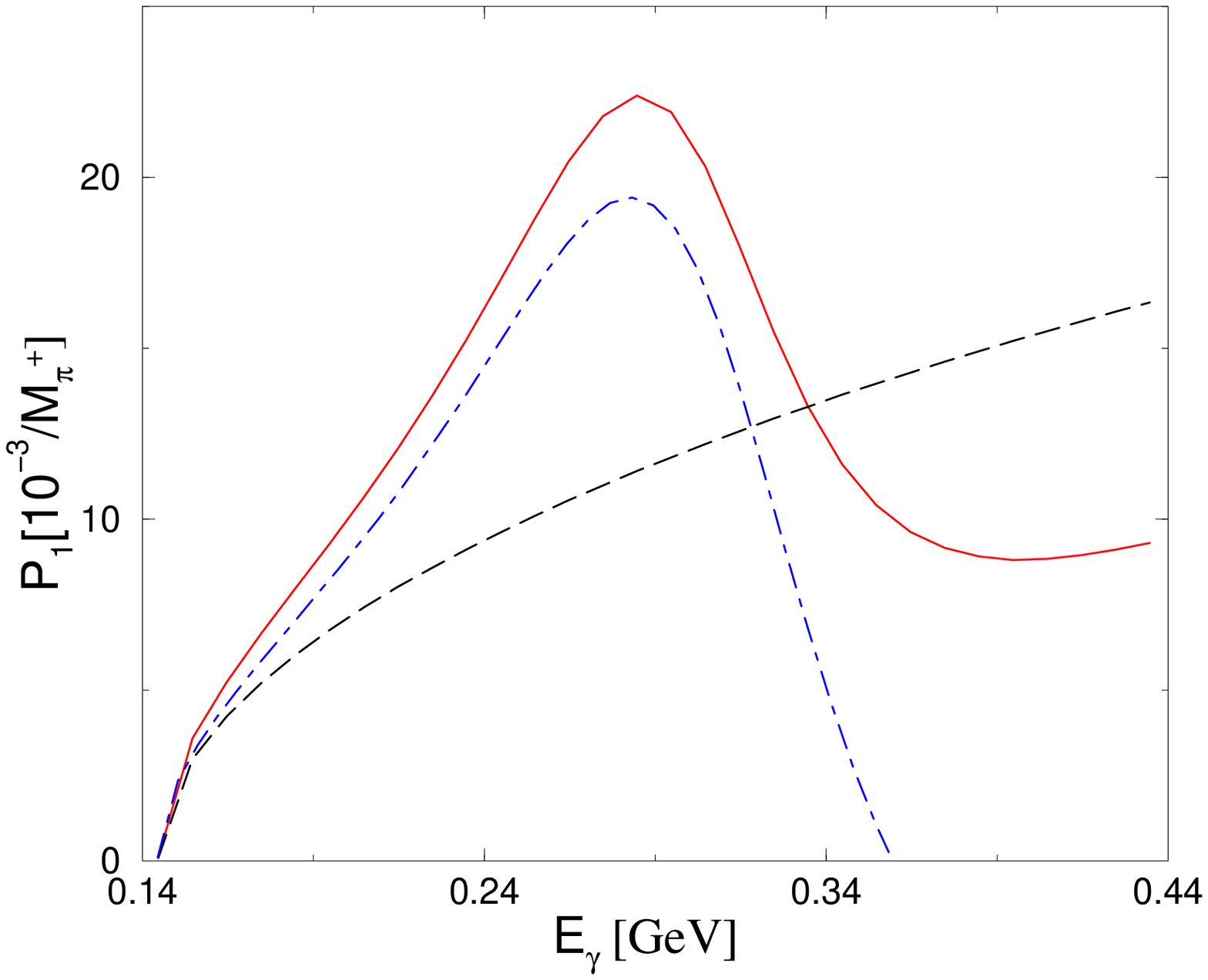,width=5cm,angle=0}
\psfig{file=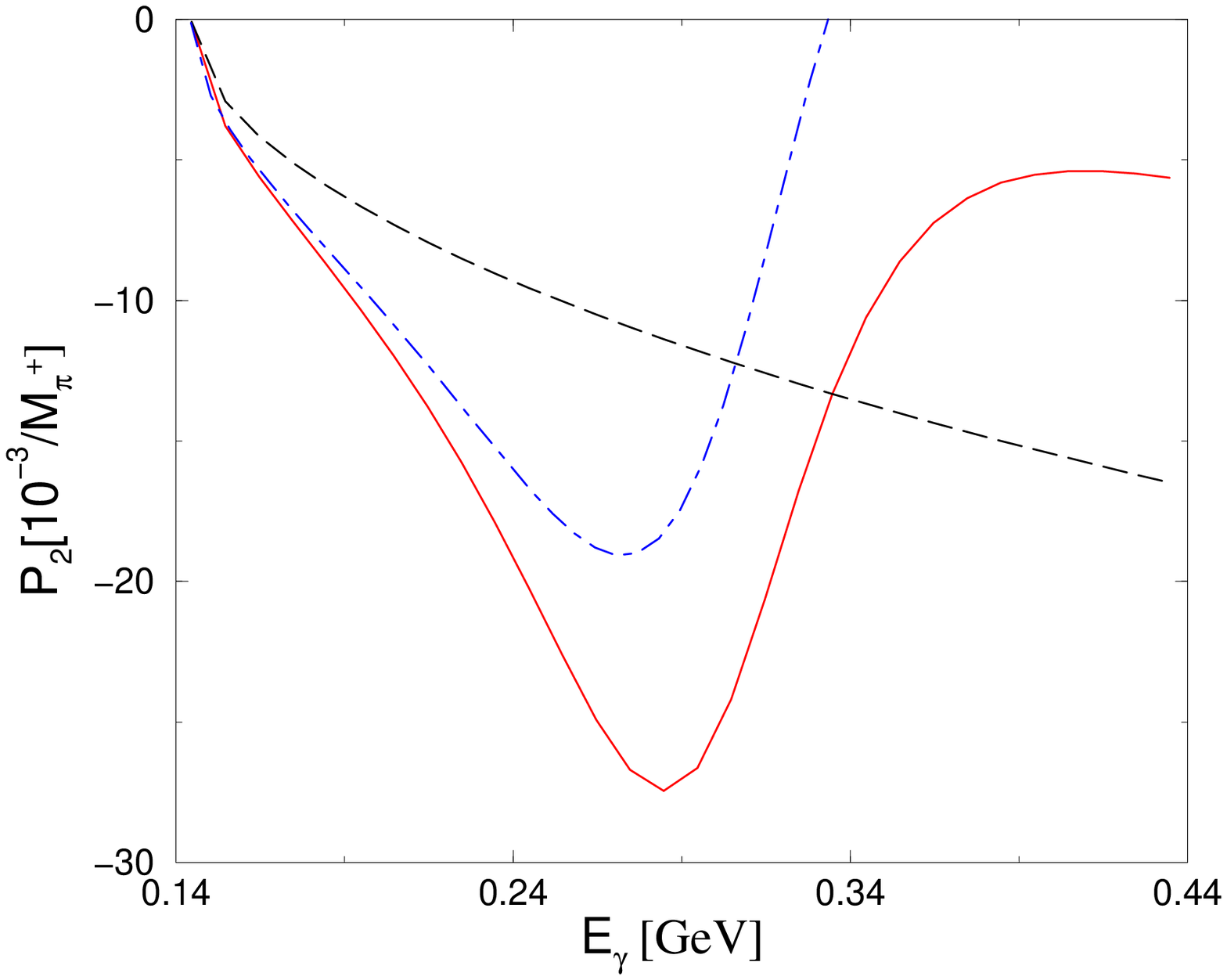,width=5cm,angle=0}
\psfig{file=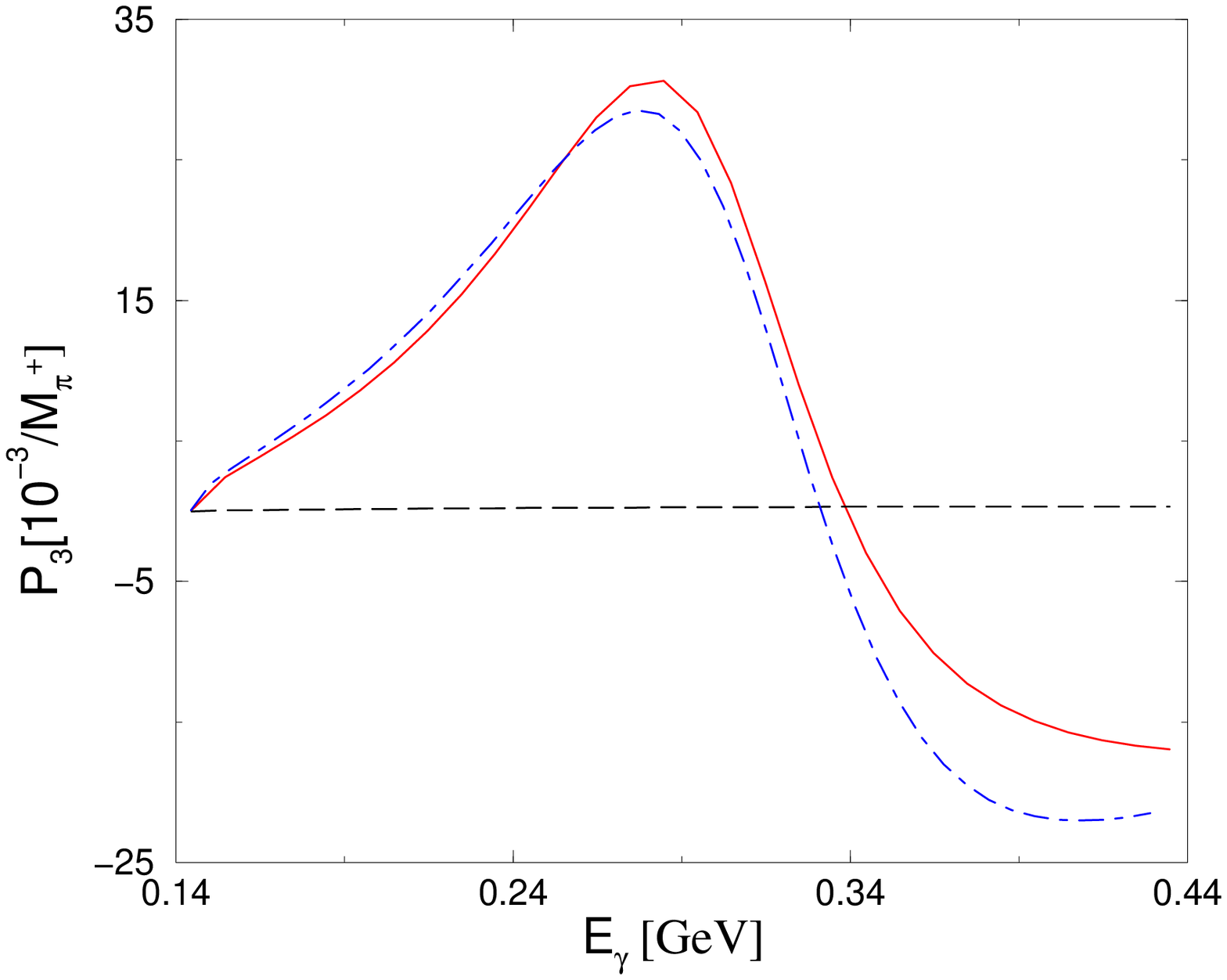,width=5cm,angle=0}
}
\vspace{-0.2cm}
\caption[fig4]{
\label{fig4} 
P-wave multipoles $P_1$, $P_2$ and $P_3$ in $\gamma p \to \pi^0 p$
as a function of the photon energy $E_\gamma$. Solid lines:
Second order calculation in the covariant SSE; dot-dashed lines: second order
calculation in covariant CHPT (utilizing infrared regularization); 
dashed line: result of the MAID analysis.
}
\end{figure}

The last topic I wish to address is the extension of these calculations to the
delta region. For that, one must include the $\Delta$ as an active degree of freedom
in the EFT. Of particular interest are the P-wave multipoles $P_i$ ($i=1,2,3$),
\begin{equation}
P_1 = 3E_{1+} + M_{1+} - M_{1-}~, \quad
P_2 = 3E_{1+} - M_{1+} + M_{1-}~, \quad
P_3 = 2M_{1+} - M_{1-}~,
\end{equation}
in terms of the more common electric ($E_{1+}$) and magnetic ($M_{1\pm}$)
P-waves. At threshold, one
can derive {\em low-energy theorems (LETs)} for the slope of $P_1$ and of $P_2$, 
while $P_3$ is dominated by the delta \cite{BKMZ}. These LETs were successfully
tested \cite{Schmidt}. In Fig.~\ref{fig4} I show the preliminary results of
a second order study in the covariant SSE \cite{BHMphoto} in comparison to the
second order covariant nucleon CHPT and the phenomenological 
MAID analysis \cite{MAID}. The description of the multipole $P_3$ is already 
quite accurate at this order, consistent with the $\Delta$-saturation of the
corresponding third order LEC in the deltaless theory (which gives a vanishing
$P_3$ at this order). For the other two multipoles loop and higher order tree
effects are needed to obtain a precise description, again consistent with the
expectations from heavy baryon CHPT \cite{BKMZ}. Still, the second order covariant SSE
calculation already captures the trend of the data (given by the MAID result).

\section{QUARK MASS DEPENDENCE OF THE NUCLEAR FORCES}
\begin{figure}[tb]
\centerline{
\psfig{file=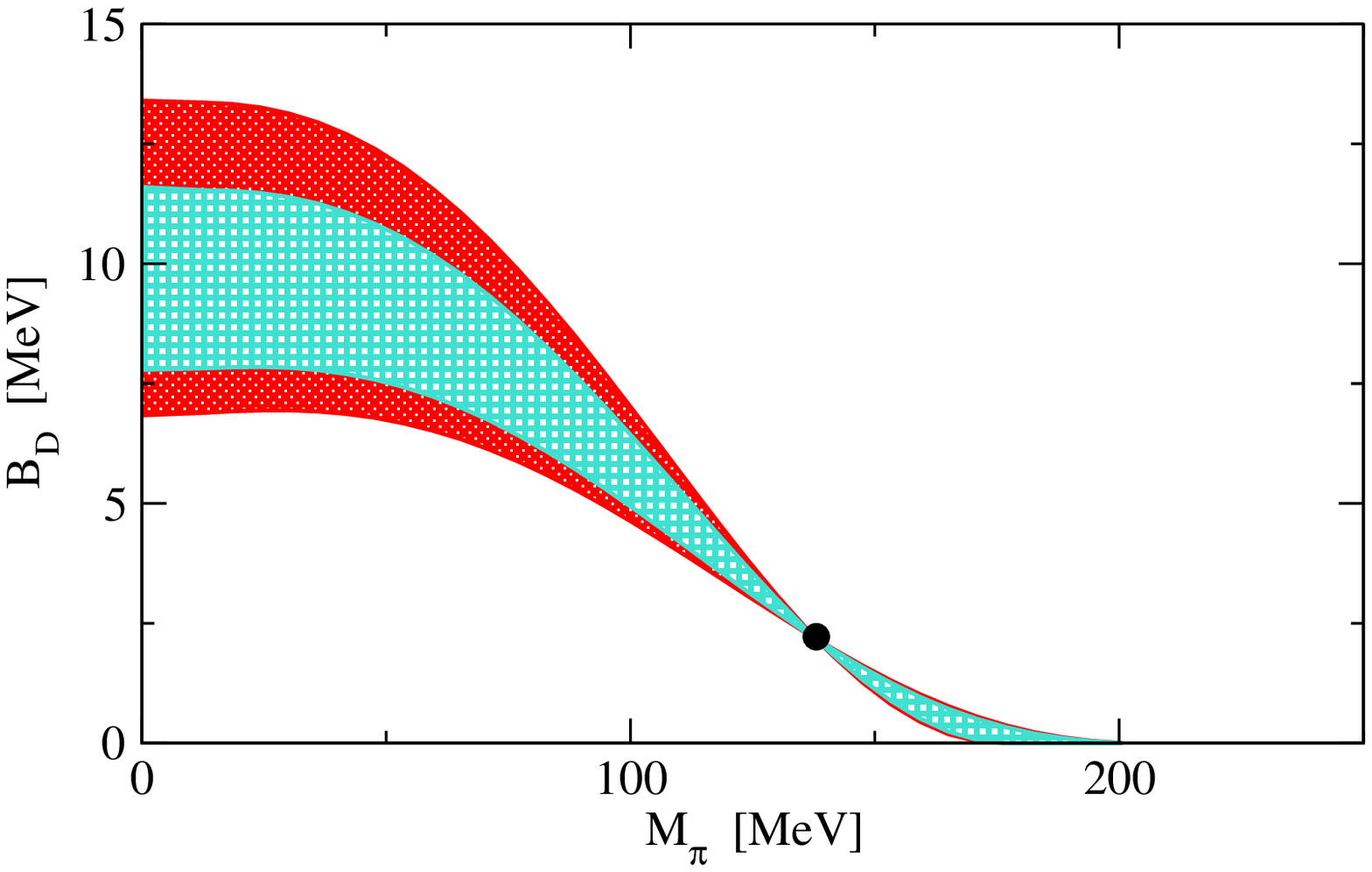,width=9.cm}
\psfig{file=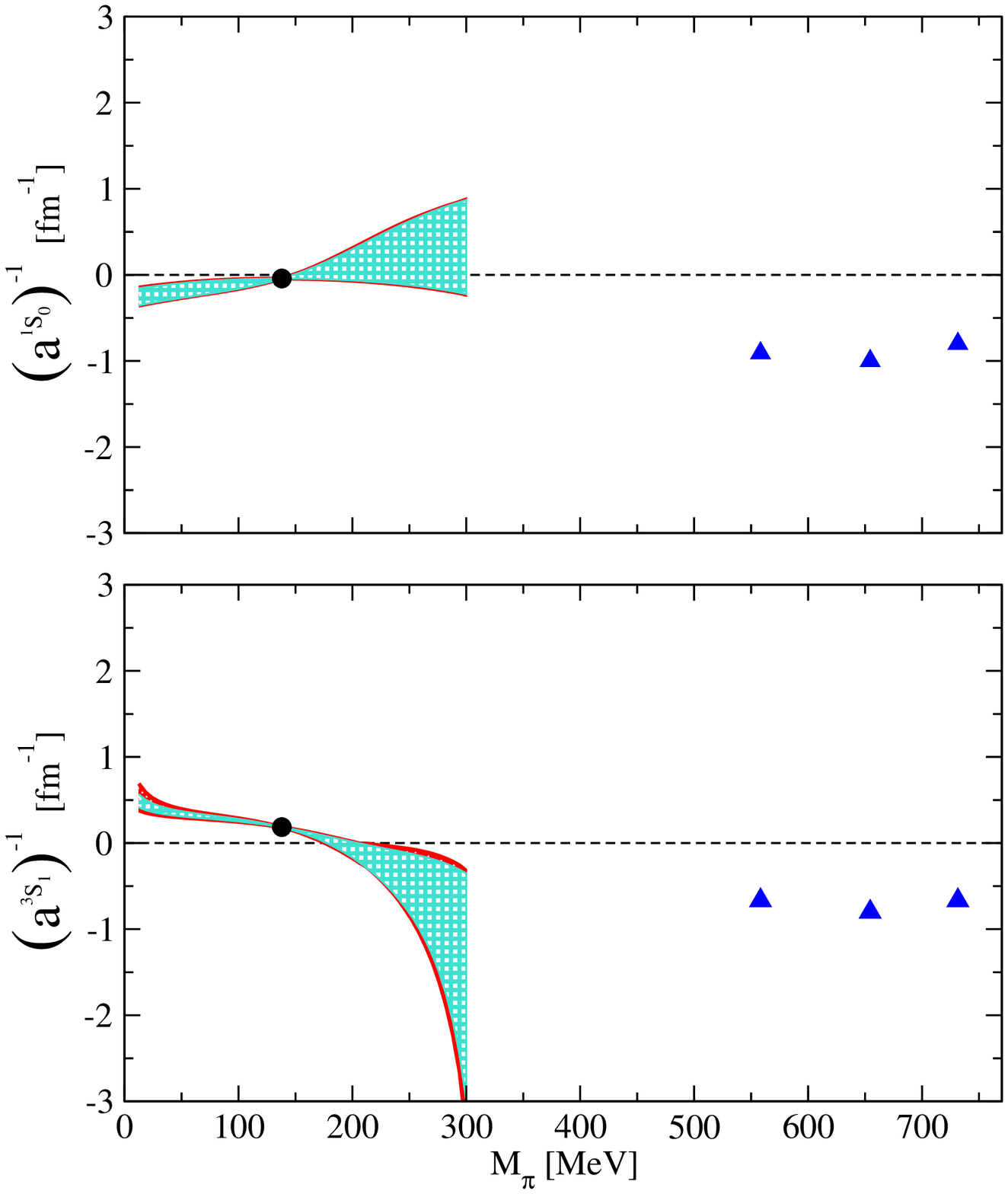,width=5.4cm}
}
\vskip -0.7 true cm
\centerline{
\parbox{1.00\textwidth}{
\caption[figcl]{
\label{figcl}
Left panel: Deuteron BE versus the pion mass. The shaded areas show 
allowed values. The light shaded band corresponds to our main result with  
the uncertainty due to the unknown LECs $\bar D_{S,T}$.
The dark shaded band gives the additional uncertainty due to the 
uncertainty of $\bar d_{16}$.
The heavy dot shows the BE for the physical case $\tilde M_\pi = M_\pi$
Right panel: The inverse S--wave scattering lengths as functions of $\tilde M_\pi$.
The shaded areas represent the allowed values according to our analysis.
The heavy dots corresponds to the values in the real world. The triangles
refer to lattice QCD results from \cite{Fukug95}.
}}}
\vspace{-0.2cm}
\end{figure}

Because of the smallness of the up and down quark masses, one does not expect
significant changes in systems of pions or pions and one nucleon when the
quark masses are set to zero (with the exception of well understood chiral
singularities like e.g. in the pion radius or the nucleon
polarizabilities). The situation is more complicated for systems of two (or
more) nucleons. Here, I report on
some work \cite{EMG} that is mostly concerned with the properties of the
deuteron and the S-wave scattering lengths as a function of the quark (pion)
mass. These questions are not only of academic interest, but also of practical
use for interpolating results from lattice gauge theory. E.g. the S-wave
scattering lengths have been calculated on the lattice using the quenched 
approximation \cite{Fukug95}. Another interesting application is related to
imposing bounds on the time-dependence of some fundamental coupling constants
from the NN sector, as discussed in \cite{Beane02}. To address this issue,
at NLO the following contributions have to be
accounted for (in addition to the LO OPE and contact terms without derivatives): 
i) contact terms with two derivatives or one $M_\pi^2$--insertion, 
ii) renormalization of the OPE,
iii) renormalization of the contact terms, and
iv) two--pion exchange (TPE).
This induces {\em explicit} and  {\em implicit} quark mass dependences. In the
first category are the pion propagator that becomes Coulomb-like in the
chiral limit or the $M_\pi^2$ corrections to the leading contact terms. These
are parameterized by the LECs $\bar D_{S,T}$ at NLO. These LECs can at present 
only be estimated using dimensional analysis and resonance saturation \cite{EMGE}. 
The implicit pion mass dependence enters at NLO through the pion--nucleon
coupling constant (note that the quark mass dependence of the nucleon
mass only enters at NNLO) expressed through the pion mass dependence of $g_A/F_\pi$
in terms of the quantity
\beq
\label{deltaCL}
\Delta = \left( \frac{g_A^2 }{16 \pi^2 F_\pi^2} - \frac{4 }{g_A}
\bar{d}_{16} + \frac{1}{16 \pi^2 F_\pi^2} \bar{l}_4 \right) 
(M_\pi^2 - \tilde M_\pi^2) - 
\frac{g_A^2 \tilde M_\pi^2}{4 \pi^2 F_\pi^2} \ln \frac{\tilde M_\pi}{M_\pi} \, .
\eeq
Here $\bar l_4$ and $\bar d_{16}$ are LECs related to 
pion and pion--nucleon interactions, and the value of the 
varying pion mass is denoted
by $\tilde M_\pi$ in order to distinguish it from the physical one denoted by
$M_\pi$. In particular, $\bar d_{16}$ has been determined in various fits to
describe $\pi N \to \pi\pi N$ data, see \cite{FBM}.
The deuteron BE as a function of the pion mass is shown in Fig.\ref{figcl},
we find that the deuteron is stronger bound in the chiral limit (CL) than in the
real world,
$B_{\rm D}^{\rm CL} =  9.6 \pm 1.9 {{+ 1.8} \atop  {-1.0}}~{\rm  MeV}$,
where the   first indicated error refers to the uncertainty in the value 
of $\bar D_{^3S_1}$ and $\bar d_{16}$ being set to its average value 
while the second indicated error shows the additional uncertainty due 
to the uncertainty in the determination of $\bar d_{16}$.
\noindent We find no other bound states, although the higher $S=1$ partial waves
rise linear with momentum due to the Coulomb-like pion propagator.
Last but not least,
we found smaller (in magnitude) and more natural values for the two 
S--wave scattering lengths in the chiral limit,
$
a_{\rm CL} (^1S_0) = -4.1 \pm 1.6 {{+ 0.0} \atop 
{-0.4}} \,{\rm fm}, {\rm   and}\,\, 
a_{\rm CL} (^3S_1) = 1.5 \pm 0.4 {{+ 0.2} \atop \\ 
{ -0.3}}\, {\rm fm}\, .
$
As stressed in \cite{EMG}, one needs lattice data for pion masses below
300 MeV to perform a stable interpolation to the physical value of $M_\pi$,
cf the right panel in Fig.~\ref{figcl}.
We conclude that nuclear physics in the chiral limit is much more natural than 
in the real world. 

\section{AN INFRARED LIMIT CYCLE IN QCD?}

Another interesting application of the quark mass dependence of the
nuclear forces is the recently conjectured infrared renormalization group
limit cycle in the three--nucleon system \cite{BH}. A limit cycle is
a non-trivial behaviour of a system under renormalization group (RG)
transformations, more precisely a closed one-dimensional orbit under RG flow
\cite{Wilson}. One of the signatures of such a limit cycle is discrete 
scale invariance, that is symmetry with respect to a scaling factor $\lambda$
of the form $\lambda^n$, with $n$ an integer. In the pionless effective
field theory, i.e. the EFT with contact interactions only, an ultraviolet
limit cycle was found for bosons with large scattering length \cite{BHvK1}
and for nucleons \cite{BHvK2}. This EFT framework embodies the {\em Efimov
effect}, namely that for systems with a very large S-wave two-particle scattering
length, there is a large number of shallow 3-body bound states with the ratio
of successive binding energies rapidly approaching a universal constant
$\lambda_0^2 = {\rm e}^{2\pi/s_0} \simeq 515$, 
with $s_0 = 1.00624$ a transcendental number \cite{Efimov}. 
For an excellent review of all the facets of such Efimov-type physics the
reader is referred to \cite{BHrev}. In the nuclear physics
case, the spin-singlet ($a (^1S_0)$) and the spin-triplet ($a (^3S_1)$)
scattering lengths are both much
larger than the range of the nuclear force $\sim 1/M_\pi$. Thus, one can describe
few-nucleon systems based on point-like interactions with a leading order three-body
force, that manifestly shows the asymptotic discrete scaling symmetry with
the scaling factor $\lambda_0 = 22.7$ and the corresponding Efimov states. Because
of the Efimov effect, the renormalization of the three-nucleon force is nontrivial
\cite{BHvK2} and involves an ultraviolet limit cycle. It is also interesting to
consider the three-nucleon system at different quark masses. As can be seen from the
right panel of Fig.~\ref{figcl}, the deuteron becomes unbound at a critical value
in the range $170\,$MeV$<M_\pi<$210$\,$MeV. In the same range, $a (^1S_0)$ also diverges
and the spin-singlet deuteron becomes bound for pion masses above 150~MeV. 
What does this imply e.g. for the triton? An exact
limit cycle would require $1/a(^1S_0) = 1/a(^3S_1) =0$. It was shown in 
\cite{BH} that for a pion mass of 175~MeV, the binding momentum of 
the $pnn$ bound state is very small and an excited state of the triton 
appears (here, one has approximated the triton binding energy by its value at the
physical pion mass). This can be considered as a hint that the
system is close to a limit cycle. In fact, to leading order in QCD, one can only
tune $M_\pi^2 \sim (m_u + m_d)$, so it was conjectured in Ref.~\cite{BH} that
by separately tuning $m_u$ and $m_d$, one could make the singlet and the triplet
scattering length diverge simultaneously. At this critical point, the deuteron and the
spin-singlet deuteron would both have zero binding energy and the triton should have
infinitely many shallow bound states, with their ratio rapidly approaching the constant
$\lambda_0^2 \simeq 515$. It is a challenge for lattice gauge theory combined with
EFT methods to indeed demonstrate the existence of such an infrared limit cycle
in QCD.

\section{BRIEF SUMMARY AND OUTLOOK}

Let me briefly summarize. I have shown that baryon chiral perturbation theory
has matured in the up and down quark sector, in particular, there exist now
covariant formulations to include matter fields with spin-1/2 and spin-3/2
(in the latter case if one considers the $N\Delta$ mass splitting as a small 
parameter). These formulations do not render the heavy baryon formalism (HBCHPT) 
obsolete but rather extend and include it. In many cases, the computational 
simplicity of
HBCHPT can and should still be used, but if one e.g. wants to make use of the
analyticity properties by utilizing dispersion relations, relativistic formulations
are the ones to be used, see e.g. the discussions in Refs.~\cite{MO,BL2,DPT,PHV}. 
Chiral extrapolation functions based on CHPT with a {\em small (moderate)} 
theoretical uncertainty can  be constructed for pion masses below 400~(550)~MeV.
Also, the extension to few-nucleon systems can be performed to a high accuracy,
for a recent precise calculation in the two-nucleon system see \cite{EGM3}. 
May be the most tantalizing 
result is the conjecture of an infrared limit cycle in QCD as described above.
I have discussed many applications of such schemes, ranging from the quark mass
dependence of the nucleon and the delta mass to double neutral pion photoproduction
off protons and more.  The extension to the strange quark sector certainly
requires more work, making proper use of unitarization methods properly matched
to CHPT amplitudes (for recent reviews, see \cite{UGM,BB}). 
Still, some results as the e.g. the SU(3) calculation of the
quark mass dependence of the nucleon mass presented here are encouraging
to further invest more effort in these topics.

\section*{ACKNOWLEDGMENTS}

It is a pleasure to thank my collaborators on these topics, V\'eronique Bernard,
Evgeny Epelbaum, Matthias Frink, Walter Gl\"ockle, Thomas Hemmert, 
Norbert Kaiser, Hermann Krebs and  
Ilka Scheller. Inspiring discussions with J\"urg Gasser and Hans-Werner Hammer
are also acknowledged. I also would like to thank the organizers
for the invitation and the excellent organization and Martin Kotulla for
supplying me with Fig.~2.
This work was supported in parts by funds provided from the DFG to the SFB/TR 16
``Subnuclear Structure of Matter'' at Bonn University.

\end{document}